\documentclass[notitlepage,nofootinbib,preprintnumbers,aps,superscriptaddress,twocolumn]{revtex4-1}
\usepackage{amsmath,amssymb,natbib,bm,color}
\usepackage{appendix}
\usepackage{graphicx}
\usepackage{slashed}
\usepackage{fontawesome5}
\definecolor{linkblue}{rgb}{0,0,0.8}
\definecolor{linkgreen}{rgb}{0,0.5,0}
\definecolor{linkdarkorange}{rgb}{1, 0.34, .2}
\definecolor{linkblack}{rgb}{0, 0, 0}
\usepackage[colorlinks,bookmarks]{hyperref}
\hypersetup{pdfpagemode=UseNone, pdfstartview=FitH, linkcolor=linkblack, citecolor=linkblack, urlcolor=linkblack}

\usepackage{verbatim}

\begin{document}
\preprint{\hbox{UTWI-16-2024, NORDITA-2024-016}}

\title{Minimal Dark Matter Freeze-in with Low Reheating Temperatures \\ and Implications for Direct Detection}

\author{Kimberly K.~Boddy}
\email{kboddy@physics.utexas.edu}
\affiliation{Texas Center for Cosmology and Astroparticle Physics, Weinberg Institute for Theoretical Physics, Department of Physics, The University of Texas at Austin, Austin, TX 78712, USA}
\author{Katherine Freese}
\email{ktfreese@utexas.edu}
\affiliation{Texas Center for Cosmology and Astroparticle Physics, Weinberg Institute for Theoretical Physics, Department of Physics, The University of Texas at Austin, Austin, TX 78712, USA}
\affiliation{The Oskar Klein Centre, Department of Physics, Stockholm University, AlbaNova, SE-10691 Stockholm, Sweden}
\affiliation{Nordic Institute for Theoretical Physics (NORDITA), 106 91 Stockholm, Sweden}
\author{Gabriele Montefalcone}
\email{montefalcone@utexas.edu}
\affiliation{Texas Center for Cosmology and Astroparticle Physics, Weinberg Institute for Theoretical Physics, Department of Physics, The University of Texas at Austin, Austin, TX 78712, USA}
\author{Barmak Shams Es Haghi}
\email{shams@austin.utexas.edu}
\affiliation{Texas Center for Cosmology and Astroparticle Physics, Weinberg Institute for Theoretical Physics, Department of Physics, The University of Texas at Austin, Austin, TX 78712, USA}

\begin{abstract}
We investigate the influence of the reheating temperature of the visible sector on the freeze-in dark matter (DM) benchmark model for direct detection experiments, where DM production is mediated by an ultralight dark photon. Here we consider a new regime for this benchmark: we take the initial temperature of the thermal Standard Model (SM) bath to be below the DM mass. The production rate from the SM bath is drastically reduced due to Boltzmann suppression, necessitating a significant increase in the portal coupling between DM and the SM to match the observed relic DM abundance. This enhancement in coupling strength increases the predicted DM-electron scattering cross section, making freeze-in DM more accessible to current direct detection experiments.
\end{abstract}
 
\maketitle

\textbf{\textit{Introduction.---}} Dark matter (DM) constitutes roughly 84\% of  the matter content in the universe~\cite{Planck:2018vyg}. While its fundamental nature remains elusive, there are continued efforts to understand DM by searching for its possible non-gravitational interactions with Standard Model (SM) particles. In particular, direct detection experiments have attained exquisite sensitivity to nuclear recoils produced from DM scattering~\cite{PandaX-II:2017hlx,XENON:2018voc,Hambye:2018dpi,LZ:2022lsv}, and more recent experiments extend their reach to sub-GeV DM masses by searching for electronic recoils~\cite{PandaX-II:2017hlx,PandaX:2022xqx,DAMIC-M:2023gxo,DAMIC-M:2023hgj,SENSEI:2020dpa,DarkSide:2022knj,XENON:2018voc}.

DM interactions with the SM may also play a key role in understanding its observed abundance throughout the universe. If DM is thermalized with the SM bath in the very early universe, its relic abundance is determined by the standard freeze-out mechanism. However, thermal production of MeV-scale DM is disallowed as it would spoil the successes of big bang nucleosynthesis (BBN)~\cite{Boehm:2013jpa,Nollett:2013pwa,Steigman:2014pfa,An:2022sva}.

Alternatively, the DM relic abundance may be generated through a freeze-in mechanism, in which the SM bath produces DM gradually through very feeble interactions, and DM never thermalizes with the SM~\cite{Hall:2009bx}. For IR freeze-in models, DM interacts with the SM via renormalizable operators. Most previous studies have assumed the process of populating the SM bath starts at high reheating temperatures $T_{\rm rh}$ well above the DM mass $m_\chi$, with the DM production rate peaking when the SM temperature drops to $T\sim m_\chi$. A popular benchmark model often referenced by the direct detection community is one in which DM freezes in via coupling to an ultralight dark photon that kinetically mixes with the SM photon~\cite{Hall:2009bx,Chu:2011be,Essig:2011nj,Chang:2019xva}. For a given DM mass, one specific value of the kinetic mixing parameter (and hence of the DM-electron scattering cross section) corresponds to production of the observed DM abundance today. This coupling must be very small to obtain the correct DM abundance, and the standard freeze-in benchmark lies at DM-electron cross sections at least an order of magnitude below current bounds.

In this \textit{letter}, we consider a different regime for DM  freeze-in production: the key ingredient is that the reheating temperature is below the DM mass, $T_{\rm rh} < m_\chi$. In this case, only particles from the high-energy tail of the SM bath velocity distribution can contribute to DM production, and the production rate  becomes exponentially suppressed regardless of the strength of its interactions with the bath. To counteract the Boltzmann suppression, matching the observed DM relic abundance requires a larger coupling of the DM sector to the SM.  This larger coupling leads to much larger interaction rates in direct detection experiments that can be probed in the near future.

Specifically, we revisit the IR freeze-in benchmark model of a kinetically mixed ultralight dark photon by exploring the effect of the reheating temperature on its abundance and the consequences for direct detection. In addition to the SM, the particle content of our model consists of the DM particle (a new dark fermion) and a dark photon with a small kinetic coupling to the SM. We treat the reheating temperature as a free parameter, allowing it to acquire values below the DM mass, but to be at least $\sim 5\, {\rm MeV}$ for consistency with the successful predictions of BBN~\cite{Hasegawa:2019jsa,Kawasaki:2000en}. Since the abundance of DM is Boltzmann suppressed for $T_{\rm rh} < m_\chi$, matching the observed DM relic abundance requires larger values of the kinetic mixing parameter and is governed by the low reheating temperature, rather than the DM mass. We find that this simple extension of the benchmark model opens up a large region of parameter space that can be probed by current direct detection experiments, and we achieve this result without resorting to more complicated dark sectors that introduce new dark degrees of freedom~\cite{Bhattiprolu:2023akk}, or a sizable initial light mediator population~\cite{Fernandez:2021iti}.

In previous work, a Boltzmann-suppressed number density was initially proposed as a solution to overproduction of very heavy and long-lived particles~\cite{Kuzmin:1997jua}. Recently, it was reintroduced in the context of DM freeze-in through the Higgs portal~\cite{Bringmann:2021sth,Cosme:2023xpa,Cosme:2024ndc,Arcadi:2024wwg} and via a massless dark photon~\cite{Gan:2023jbs}. Direct detection of DM for low reheating temperatures has also been studied in UV freeze-in models in which DM interacts with the SM through non-renormalizable interactions~\cite{Essig:2011nj,Bhattiprolu:2022sdd}.


\textbf{\textit{General Set-Up.---}} We assume the standard \textit{kinetic mixing} portal for the production of DM from SM particles~\cite{GALISON1984279,Holdom:1985ag,Ackerman:2008kmp,Feng:2008mu,Feng:2009mn}. We consider a massive dark photon $\gamma'$ kinetically mixed with the SM hypercharge:

\begin{equation}
    \mathcal{L}\supset
    \frac{\epsilon}{2 \cos\theta_W}F^\prime_{\mu\nu}F_Y^{\mu\nu},
\end{equation}
where $F_Y^{\mu\nu}$ and $F^\prime_{\mu\nu}$ correspond to the field strengths of the SM hypercharge and the dark photon, respectively; $\theta_W$ is the Weinberg mixing angle; and $\epsilon\ll 1$ is the kinetic mixing parameter.

In the minimal scenario we consider here, the dark photon is coupled to the current of a Dirac fermionic DM particle $\chi$. Motivated by the thresholds of ongoing direct detection experiments, we explore the MeV-TeV range for the DM mass $m_\chi$. The dark photon has a non-zero but negligible mass $m_{\gamma^\prime}$, which enhances the direct detection cross section at low momentum transfer $q$~\cite{Essig:2011nj}. Without loss of generality, the mass of the dark photon is assumed to be below $\sim 10^{-15}\,$eV. In this range, the kinetic mixing is unaffected by constraints from COBE/FIRAS~\cite{Fixsen:1996nj,Caputo:2020bdy} and black hole superradiance~\cite{Baryakhtar:2017ngi,Siemonsen:2022ivj}.
Following the standard prescription~\cite{Chu:2011be}, we present our results in terms of the \textit{portal coupling} $\kappa\equiv \epsilon \sqrt{\alpha^\prime/\alpha}$, where $\alpha$ and $\alpha'$ are the electromagnetic and dark fine structure constants, respectively.

We assume the universe predominantly reheats to the visible sector at a temperature $T=T_{\rm{rh}}$ with a vanishingly small initial DM abundance~\cite{Hall:2009bx}. The evolution of the DM number density $n_\chi$ is governed by the Boltzmann equation
\begin{equation}
    \dot{n}_\chi+3H n_\chi=\sum_{B} \langle \sigma_{B\overline{B} \rightarrow \chi\overline{\chi}} v \rangle (n^{\rm eq}_\chi)^2, \label{eq:nchi}
\end{equation}

where the dot denotes a derivative with respect to cosmic time, the angle brackets denote a thermal average over relative velocities $v$, $H$ is the Hubble parameter, $n^{\rm eq}_\chi$ is the DM number density if DM were in thermal equilibrium with the SM bath, and the sum runs over all SM particles $B$ that produce $\chi \overline{\chi}$ at tree level (i.e., leptons, quarks/light charged mesons for temperatures above/below the QCD phase transition, and $W$ bosons).

Recasting Eq.~\eqref{eq:nchi} in terms of $x\equiv m_\chi/T$ and the DM yield $Y_\chi\equiv n_\chi/s$, where $s$ is the entropy density, we obtain the solution
\begin{equation}
     Y_\chi(x)=\int_{x_{\rm rh}}^x dx^\prime \frac{s}{\overline{H}x^\prime}\left[\sum_{B}\langle \sigma_{B\overline{B} \rightarrow \chi\overline{\chi}} v \rangle  (Y^{\rm eq}_\chi)^2\right], \label{eq:Ychi}
\end{equation}
where $H/\overline{H}= 1+ \frac{1}{3}\frac{d\ln g_{\star,s}}{d\ln T}$ and $g_{\star,s}$ is the effective number of relativistic degrees of freedom associated with entropy.
We impose the initial condition $Y(x_{\rm rh})\approx 0$, which reflects the assumption that the universe does not produce any substantial amount of DM during reheating.

Finally, for all the results presented in the following sections, we use the publicly available code introduced in Ref.~\cite{Bhattiprolu:2023akk}, which provided an updated (and corrected) prediction for the standard freeze-in benchmark. We have modified the code to accommodate different reheating temperatures.

\textbf{\textit{Impact of Low Reheating Temperatures.---}} In the typical scenario where the reheating temperature $T_{\rm{rh}}$ is above the DM mass, the precise value $T_{\rm{rh}}$ is not important when evaluating the DM relic abundance: all the processes connecting the visible and dark sectors are IR-dominated, such that the bulk of the DM production occurs at lower temperatures corresponding to $x\sim \mathcal{O}(1)$, and the final abundance is independent of initial conditions. Thus, we can simply set $x_{\rm{rh}}\approx 0$ for the (standard) freeze-in benchmark.

\begin{figure}[t]
    \centering
    \includegraphics[width=\linewidth]{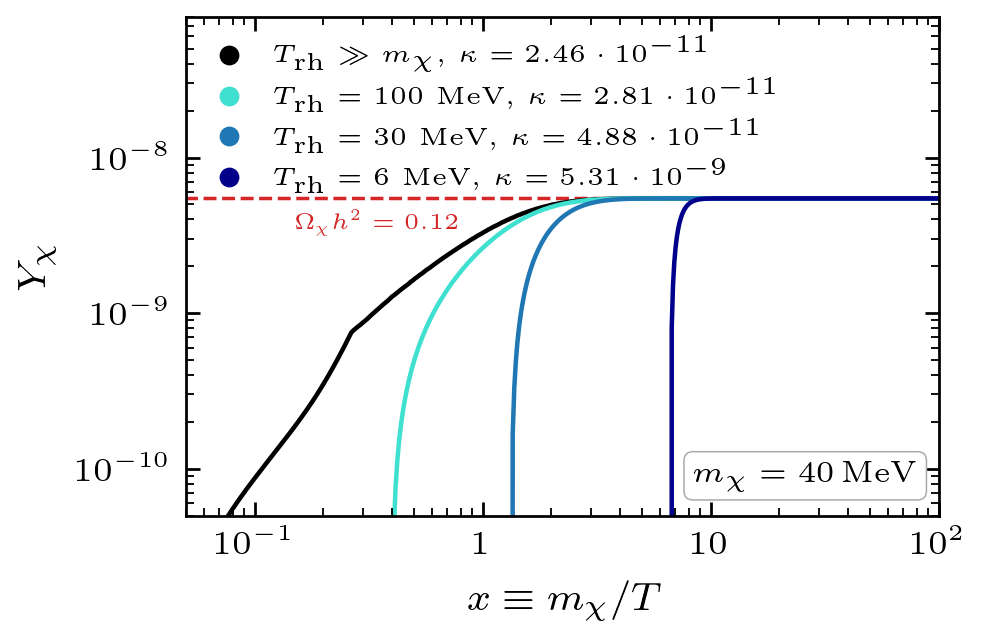}
    \caption{The evolution of the DM yield $Y_\chi \equiv n_\chi /s$ as a function of $x\equiv m_\chi/T$ for a fixed DM mass $m_\chi=40\,$MeV and reheating temperatures $T_{\rm rh}=\{6\,\mathrm{MeV},\,\,30\,\mathrm{MeV},\,\,100\,\mathrm{MeV}\}$ (shown in navy, blue, and cyan). The standard case with $T_{\rm rh}\gg m_\chi$ is shown in black. The yield corresponding to the observed relic abundance is shown as the dashed orange line.}
    \label{fig:0}
\end{figure}

However, the freeze-in prediction is insensitive to initial conditions only as long as the reheating temperature of the SM bath is much higher than the DM mass, i.e.\ for $x_{\rm rh}\ll 1$~\cite{Cosme:2023xpa,Cosme:2024ndc}. In fact, if $x_{\rm rh}\sim 1$ and especially when $x_{\rm rh}\gg 1$, DM production is exponentially suppressed, as only SM particles in the tail of their velocity distributions have enough energy to annihilate into DM particles that have a mass $m_\chi\gg T$. In this scenario, the final DM abundance is quickly established and becomes very sensitive to the reheating temperature of the SM. In this sense, the DM production process is still IR-dominated, but the lowest relevant energy scale is the reheating temperature itself rather than the DM mass. Importantly, to counteract the suppressed production rate in these low reheating scenarios, a larger coupling with the SM is required in order to obtain the observed DM abundance today.

We present the full evolution of the DM yield for different values of the reheating temperature in Fig.~\ref{fig:0}, and we match to the observed DM relic abundance of $\Omega_{\rm CDM}h^2=0.12$~\cite{Planck:2018vyg}. For illustration, we fix $m_\chi=40\,$MeV and show the standard evolution for $T_{\rm{rh}}\gg m_\chi$, as well as the cases with $T_{\rm rh}=\{6\,\mathrm{MeV},\,\,30\,\mathrm{MeV},\,\,100\,\mathrm{MeV}\}$. BBN constrains the minimum reheating temperature~\cite{Hasegawa:2019jsa,Kawasaki:2000en}, which we safely take to be $T_{\rm rh}=6\,$MeV in this work. As expected, for reheating temperatures significantly below the DM mass, we observe that the DM yield freezes in very quickly to its final value and requires a larger value of the portal coupling $\kappa$. For the example shown in Fig.~\ref{fig:0}, we have $\kappa(T_{\rm rh}=6\,\mathrm{MeV})/\kappa(T_{\rm rh}\gg m_\chi)\sim 10^{2}$.

We can understand the required enhancement in $\kappa$ analytically by solving Eq.~\eqref{eq:Ychi} in the limit $T_{\rm rh}\ll m_\chi$ and comparing the obtained relic abundance to the conventional result for $T_{\rm{rh}}\gg m_\chi$. We find that an enhancement of
\begin{equation}
    \frac{\kappa(T_{\rm rh}\ll m_\chi)}{\kappa (T_{\rm rh}\gg m_\chi)}\sim \sqrt{ x_{\rm rh}}e^{x_{\rm{rh}}} \label{eq:kappa}
\end{equation}
over the standard freeze-in prediction is needed.

\begin{figure}[t]
    \centering
    \includegraphics[width=\linewidth]{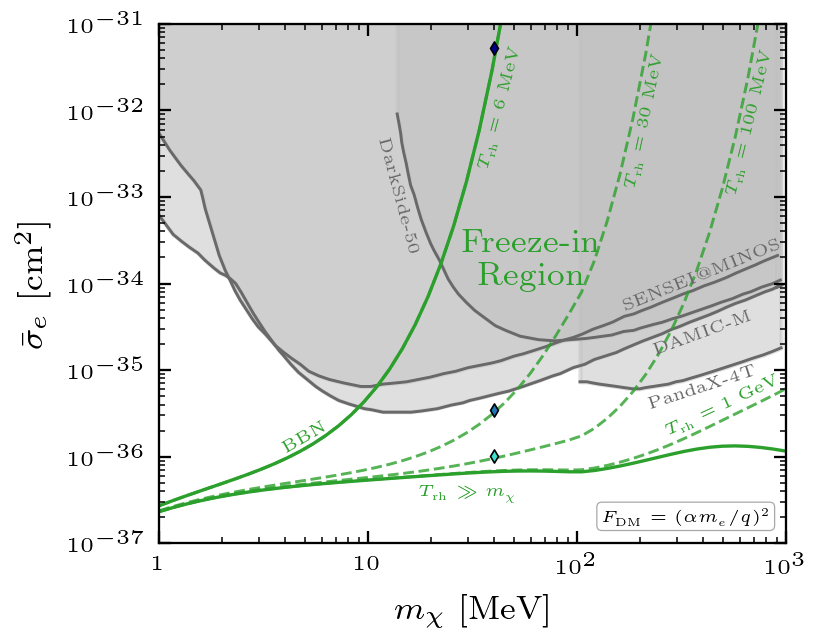}
    \caption{The updated range of direct detection cross sections allowed by the freeze-in scenario (labeled ``Freeze-in Region''), when accounting for all possible reheating temperatures above the BBN bound. Only for $T_{\rm{rh}}\gg m_\chi$ do we recover the previously understood freeze-in benchmark, which corresponds to the bottom solid green line. Shaded in gray are the current experimental constraints from DAMIC-M~\cite{DAMIC-M:2023gxo,DAMIC-M:2023hgj}, DarkSide~\cite{DarkSide:2022knj}, PandaX~\cite{PandaX:2022xqx}, SENSEI~\cite{SENSEI:2020dpa} and XENON1T~\cite{XENON:2018voc,Hambye:2018dpi}. }
    \label{fig:1}
\end{figure}

\textbf{\textit{Implications for Direct Detection.---}} The enhancement in Eq.~\eqref{eq:kappa} emphasizes the potentially enormous impact that a low reheating temperature has on the size of the portal coupling $\kappa$ associated with freeze-in DM. In turn, the corresponding direct detection rates in these low-reheating scenarios far surpass the standard freeze-in benchmark.

\begin{figure}[t]
    \centering
    \includegraphics[width=\linewidth]{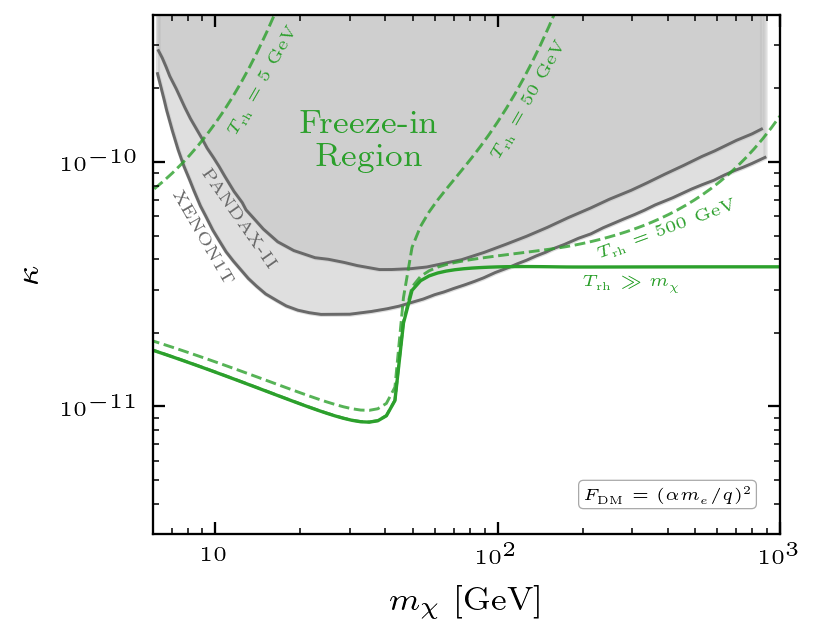}
    \caption{Updated range of portal couplings $\kappa$ allowed by the freeze-in scenario (labeled ``Freeze-in Region") when accounting for reheating temperatures $T_{\rm{rh}}\lesssim m_\chi$. The previously understood freeze-in benchmark with $T_{\rm{rh}}\gg m_\chi$ corresponds to the bottom solid green line. Shaded in gray are the current experimental constraints from PandaX~\cite{PandaX-II:2017hlx} and XENON1T~\cite{XENON:2018voc,Hambye:2018dpi}. }
    \label{fig:2}
\end{figure}

The reference DM-electron cross section used in direct detection literature is~\cite{Essig:2015cda}
\begin{equation}
    \overline{\sigma}_e=\frac{16\pi \mu_{\chi e}^2\alpha^2\kappa^2}{(\alpha m_e)^4},
\end{equation}
where $m_e$ is the electron mass, and $\mu_{\chi e}$ is the DM-electron reduced mass. For our case of interest, the interaction is mediated by an ultralight dark photon, and the DM-electron cross section is $\overline{\sigma}_e$ multiplied by the square of the form factor $F_\mathrm{DM} = (\alpha m_e / q)^2$.

In Fig.~\ref{fig:1}, we illustrate the updated range of direct detection cross sections allowed by the freeze-in scenario for a DM mass in the MeV-GeV range, when accounting for all possible reheating temperatures above the minimum reheating temperature allowed by BBN. For $T_{\rm rh}\gg m_\chi$, we recover the standard freeze-in benchmark. As the reheating temperature decreases, the freeze-in prediction moves to larger cross sections, specifically for values of the DM mass below the reheating temperature under consideration. We show the freeze-in benchmarks for $T_{\rm{rh}}=\{30\,\mathrm{MeV},\,100\,\mathrm{MeV},\, 1\,\mathrm{GeV}\}$ as dashed green lines in the figure. As expected, departures from the standard freeze-in benchmark occur only for $m_\chi\gtrsim T_{\rm{rh}}$. Additionally, we mark the $\overline{\sigma}_e$ predictions from the examples presented in Fig.~\ref{fig:0} as diamonds in their corresponding color from the legend in Fig.~\ref{fig:0}.

We find that the freeze-in scenario is not simply confined to the bottom solid green curve in Fig.~\ref{fig:1}, but rather it encompasses the entire region between the BBN bound and standard benchmark.
In shaded gray, we also show the current electronic-recoil direct detection exclusion limits from DAMIC-M~\cite{DAMIC-M:2023gxo,DAMIC-M:2023hgj}, DarkSide~\cite{DarkSide:2022knj}, PandaX~\cite{PandaX-II:2017hlx,PandaX:2022xqx}, SENSEI~\cite{SENSEI:2020dpa} and XENON1T~\cite{XENON:2018voc,Hambye:2018dpi}.
These exclusion curves intersect the broadened freeze-in region, indicating direct detection experiments are already actively probing the sub-GeV freeze-in DM scenario with low reheating temperatures.

There is a similar story for freeze-in DM in the GeV-TeV mass range, as shown in Fig.~\ref{fig:2}. Instead of just the standard freeze-in benchmark corresponding to $T_{\rm rh}\gg m_\chi$, the freeze-in region covers all parameter space shown above the standard benchmark. For illustration, we also plot the cases for $T_{\rm{rh}}=\{5\,\mathrm{GeV},\,50\,\mathrm{GeV},\, 500\,\mathrm{GeV}\}$.
We show the current nuclear-recoil direct detection exclusion limits from PandaX~\cite{PandaX-II:2017hlx} and XENON1T~\cite{XENON:2018voc,Hambye:2018dpi}. We find that for $m_\chi\lesssim T_{\rm rh}$, the required portal coupling $\kappa$ is significantly enhanced to values that are being actively probed by current experiments.

\textbf{\textit{Discussion.---}}
In this \textit{letter}, we have explored the impact of the reheating temperature on the benchmark freeze-in model in which DM is produced from the SM particles through a kinetically mixed ultralight dark photon. We show that when the reheating temperature is below the mass of DM, its production rate from the bath is exponentially suppressed, and a larger portal coupling is required to achieve the observed relic abundance. This enhancement consequently lifts the freeze-in benchmark target for direct detection towards larger DM-electron scattering cross sections. The possible range for the reheating temperature, therefore, opens up a large region of the DM parameter space that is already being probed by current direct detection experiments. We argue that the freeze-in benchmark target should be regarded as an extended region defined by the reheating temperature, rather than a single curve.  In this sense, within the context of the minimal freeze-in scenario analyzed here, a potential future detection that lies between the current observational upper limits and the traditional freeze-in benchmark would directly probe the reheating temperature, providing important insights into the conditions of the universe in its earliest moments.

Lastly, it is worth noting that in this \textit{letter}, we assume that the maximum temperature of the bath coincides with the reheating temperature. This assumption is consistent with instantaneous reheating or reheating models that keep the reheating temperature close to the maximum temperature~\cite{Co:2020xaf,Cosme:2023xpa,Cosme:2024ndc, Barman:2024mqo}. Reheating scenarios that predict a maximum temperature larger than the reheating temperature (e.g.\ Refs.~\cite{Chung:1998rq,Giudice:2000ex,Kolb:2003ke}) are also applicable as long as the maximum temperature does not exceed the mass of DM.

\textit{\textbf{Acknowledgments.---}\label{sec:Acknowledgments}} 
We thank Gilly Elor for the valuable discussion. K.B.\ acknowledges support from the National Science Foundation under Grant No.~PHY-2112884.
K.F.\ is Jeff \& Gail Kodosky Endowed Chair in Physics at the University of Texas at Austin, and K.F.\ and G.M.\  are grateful for support via this Chair. K.F., G.M., and B.S.E.\  acknowledge support by the U.S.\ Department of Energy, Office of Science, Office of High Energy Physics program under Award Number DE-SC-0022021. K.F.\ and G.M.\ also acknowledge support from the Swedish Research Council (Contract No.~638-2013-8993).

\bibliographystyle{apsrev4-1}
\bibliography{letter}

\end{document}